\documentclass[twocolumn, english,showpacs,amsmath,amssymb]{revtex4}
\usepackage{amsmath}
\usepackage{amssymb}
\usepackage{graphicx}
\usepackage{babel}
\usepackage{times}
\usepackage{hyperref}

\allowdisplaybreaks[3]

\graphicspath{{Figures/}}

\begin{document}

\title{Weyl spin liquids}

\author{M. Hermanns}
\author{K. O'Brien}
\author{S. Trebst}
\affiliation{Institute for Theoretical Physics, University of Cologne, 50937 Cologne, Germany}

\date{\today}

\begin{abstract}
The fractionalization of quantum numbers in interacting quantum-many body systems is a central motif in condensed matter physics with 
prominent examples including the fractionalization of the electron in quantum Hall liquids or the emergence of magnetic monopoles in spin-ice materials. Here we discuss the fractionalization of magnetic moments in three-dimensional Kitaev models into Majorana fermions (and a $\mathbb Z_2$ gauge field) and their emergent collective behavior. We analytically demonstrate that the Majorana fermions form a Weyl superconductor for the Kitaev model on the recently synthesized hyperhoneycomb structure of $\beta$-Li$_2$IrO$_3$ when applying a magnetic field.
We characterize the topologically protected bulk and surface features of this state, which we dub a Weyl spin liquid, 
including thermodynamic and transport signatures.
\end{abstract}

\pacs{75.10.Kt, 03.65.Vf, 71.20.Be}
\maketitle

\begin{figure}[b]
  \includegraphics[width=.8\columnwidth]{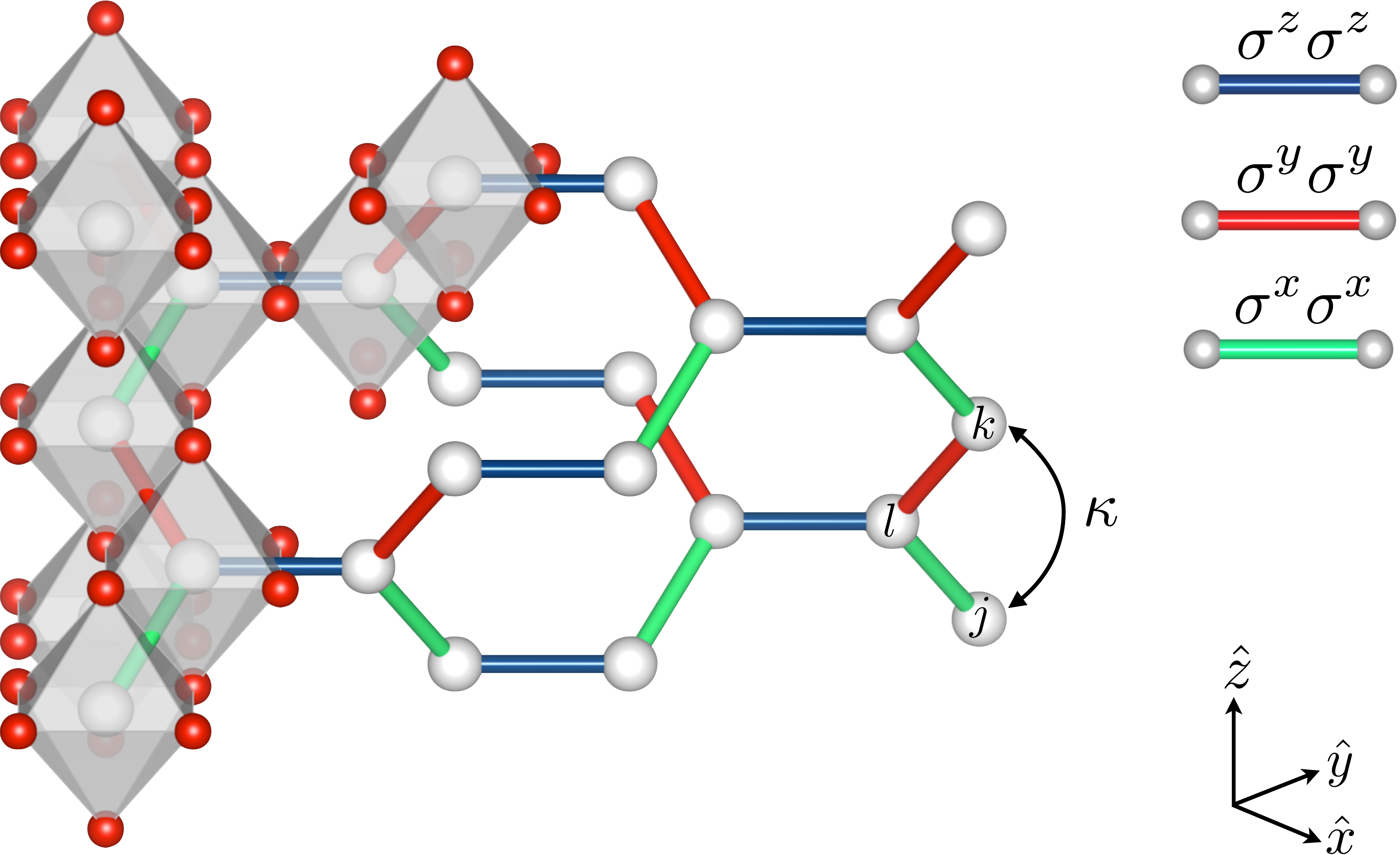}
  \caption{
    (color online) The tricoordinated hyperhoneycomb lattice synthesized for $\beta$-Li$_2$IrO$_3$.  IrO$_6$ octahedra are indicated on the left. Green, red, and blue bonds correspond to an Ising-type interaction of $\sigma^x \sigma^x$, $\sigma^y \sigma^y$, and $\sigma^z \sigma^z$ type, respectively. The three-spin interaction $\sigma^x_j \sigma^y_k \sigma_l^z$ induces a next-nearest neighbor hopping term between sites $j$ and $k$ in the effective Majorana model \eqref{eq:eff}. 
   }
  \label{Fig:Hyperhoneycomb}
\end{figure}

\noindent 
One of the most intriguing phenomena in strongly correlated systems is the fractionalization of quantum numbers, i.e. the low-temperature emergence of novel quantum numbers which are distinct from those of the original constituents of the quantum many-body system.
Familiar examples include the spin-charge separation in one-dimensional metallic systems \cite{SpinChargeSeparation}, the fractionalization of the electron in certain quantum Hall states \cite{FQH}, and the emergence of monopoles in spin ice \cite{MonopolesSpinIce} or chiral magnets \cite{MonopolesChiralMagnets}.
In this paper, we discuss the fractionalization of magnetic moments in three-dimensional generalizations of the Kitaev model \cite{Kitaev} and the collective behavior of the emergent Majorana fermionic degrees of freedom. The latter form metallic states whose precise character intimately depends on the underlying lattice structure. 
For the two-dimensional honeycomb Kitaev model it is well known that the Majorana fermions form a semimetal with two gapless {\em Dirac points} \cite{Kitaev}. Recently, three-dimensional lattice structures have been considered for which the emergent Majorana fermions form metallic states with  gapless modes either along a {\it Fermi line} \cite{Mandal09} or a two-dimensional {\it Fermi surface} akin to a conventional metal \cite{hyperoctagon}. The common feature of these lattices is that they preserve the tricoordination of the vertices familiar from the honeycomb lattice.
An example of such a three-dimensional  lattice structure is the so-called hyperhoneycomb lattice illustrated in Fig.~\ref{Fig:Hyperhoneycomb},  which has recently been synthesized for the Iridate compound Li$_2$IrO$_3$ \cite{hyperTakagi}. Like other 5d transition metal oxides, Li$_2$IrO$_3$ exhibits an intricate interplay of electronic correlations, crystal field effects, and strong spin-orbit coupling leading to the formation of a Mott insulator where the local moments are spin-orbit entangled $j=1/2$ Kramers doublets \cite{Sr2IrO4}. It has been argued \cite{Jackeli09} that the microscopic interactions between these local $j=1/2$ moments realize Kitaev-type Hamiltonians with recent experiments indeed confirming such a spatially highly anisotropic exchange \cite{hyperAnalytis,Biffin,Coldea}.

Here we demonstrate that in the presence of a magnetic field (or any other time-reversal symmetry breaking term),
the emergent Majorana fermions in such three-dimensional Kitaev models can form yet another collective state -- a Weyl superconductor. The band structure of the latter is characterized by the presence of gapless  {\em Weyl points} in the bulk and the formation of  topologically protected gapless  {\em Fermi arcs} on the surface \cite{WeylSM}.  
Keeping in mind that this physics in fact plays out in a quantum spin system, we dub this highly unconventional emergent state a Weyl spin liquid. 
The key distinction when compared to  electronic Weyl semimetals or superconductors is that it arises in an electronic Mott {\em insulator}. 
In the latter, the electronic degrees of freedom are frozen out, but the collective state of the localized moments mimics the formation of an itinerant electronic state. 
The analytical tractability of the Kitaev model allows us to comprehensively discuss the intriguing  facets of this state in the following.

 
\begin{figure}
  \includegraphics[width=\columnwidth]{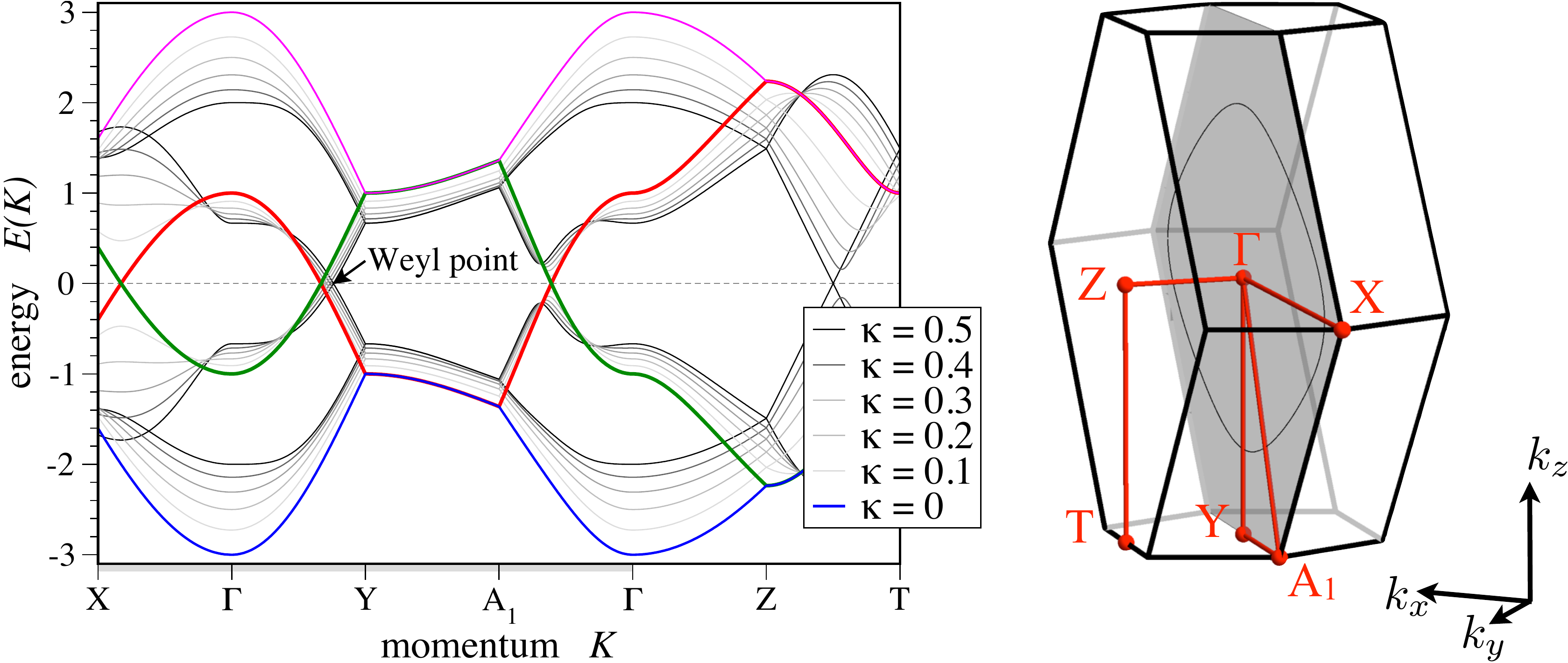}
   \caption{
    (color online)
	The energy dispersion of the hyperhoneycomb Kitaev model for various values of $\kappa$ 
	(parametrizing the effective magnetic field)
	along certain high-symmetry lines indicated in the Brillouin zone on the right hand side.
	The grey hexagon indicates the plane $k_x=-k_y$ on which the line of gapless mode
	(black line) is located.
	}  
  \label{Fig:Dispersion}
\end{figure}
 
\noindent {\it Model.--} 
Specifically, we consider a Kitaev model 
on the hyperhoneycomb lattice 
\begin{equation}
  H_{\rm Kitaev} = -J_K \sum_{\gamma \rm-bonds}  \sigma_i^\gamma \sigma_j^\gamma\,,  
  \label{eq:spinH}
\end{equation}
which favors nearest-neighbor SU(2) spin-1/2s (represented by the Pauli matrices $\sigma_i$)  to align their $x$, $y$ or $z$ components depending on the bond directions of the hyperhoneycomb lattice (as color-coded in Fig.~\ref{Fig:Hyperhoneycomb}). Remarkably, this highly frustrated spin model can be solved exactly. 
In close analogy to Kitaev's solution~\cite{Kitaev} of the two-dimensional honeycomb model, one represents the spins in terms of four Majorana fermions $\sigma_j^{\gamma} = i\, b_j^{\gamma} c_j^{\phantom\gamma}$ and regroups the two Majorana fermions associated with a bond into an operator $\hat{u}_{ij} = i\,b_i^{\gamma}b_j^{\gamma}$ 
whose $\pm 1$ eigenvalues can be identified with a static $\mathbb{Z}_2$ gauge field. 
One thereby maps the original interacting spin model to a {\em free}  fermion Hamiltonian of Majorana degrees of freedom hopping in the presence of a static $\mathbb Z_2$ gauge field.
Diagonalizing this Hamiltonian 
\cite{footnoteZeroFlux},  one readily obtains the band structure of this model with four distinct bands arising from the four sites of the unit cell of the hyperhoneycomb lattice. 
As shown in Fig.~\ref{Fig:Dispersion} the system exhibits gapless modes located along a closed loop in the $k_y=-k_x$ plane of the Brillouin zone \cite{Mandal09}. 
Close inspection \cite{Kitaev,hyperoctagon} of the Hamiltonian further reveals that the gapless modes are protected by time-reversal symmetry and indeed stable against various perturbations \cite{hyperKim,hyperKim2,hyperKim3,hyperKimchi}, as well as thermal fluctuations \cite{Nasu1,Nasu2,Nasu3}. 
Any time-reversal invariant perturbation to the Hamiltonian can only deform the line, but not immediately gap out the gapless modes (see the supplemental material).

We now ask what effects are induced by time-reversal symmetry breaking perturbations such as a magnetic field. In particular, we study a term $- \sum_{j}\vec h \cdot \vec \sigma_j$ where the magnetic field points along the 111 direction. This augmented Kitaev model is no longer exactly solvable per se. However, one can perturbatively derive a low-energy effective model that remains exactly solvable, again similar to the two-dimensional Kitaev model \cite{Kitaev}. This effective model is obtained by observing that the static  $\mathbb Z_2$ gauge field allows to perform perturbation theory in the zero-flux ground state sector as long as the strength of the magnetic field remains smaller than the flux gap, which is approximately 0.2 $J_K$ around the isotropic point. The first non-trivial contributions arise at third-order and yield an effective Hamiltonian
\begin{equation}
  H_{\rm eff} = -J_K \sum_{\gamma \rm-bonds}  \sigma_i^\gamma \sigma_j^\gamma - \tilde\kappa \sum_{j,k,l} \sigma_j^\alpha \sigma_k^\beta \sigma_l^\gamma\,,  
  \label{eq:mag}
\end{equation}
with a three-spin coupling constant $\tilde \kappa\sim h_x h_y h_z/J_K^2$. The sites $j$ and $k$ denote two distinct nearest-neighbors of site $l$ and ($\alpha$,  $\beta$, $\gamma$) is a permutation of $(x,y,z)$ with bond $jl$ ($kl$) being of type $\alpha$ ($\beta$) as illustrated in Fig.~\ref{Fig:Hyperhoneycomb}.
Recasting this Hamiltonian in terms of the Majorana fermions yields a non-interacting model of fermions hopping between nearest and next-nearest neighbors
\begin{equation}
  H_{\rm eff} = i  J_K \sum_{\langle j,k\rangle} u_{jk} c_j c_k  -i \tilde \kappa \sum_{\langle \! \langle j,k \rangle \! \rangle} \tilde u_{jk} c_j c_k \,,
  \label{eq:eff}
\end{equation}
where the coefficient of the nearest neighbor hopping is given by $u_{jk}=1 (-1) $ for $j$ on the odd (even) sublattice and the coefficient in front of the next-nearest neighbor hopping term $\tilde u_{jk}$ can be determined from the underlying three-spin interaction, $\sigma_j^\alpha \sigma_k^\beta \sigma_l^\gamma$, to be $\tilde u_{jk} =-\epsilon^{\alpha\beta\gamma}$, where $\epsilon^{\alpha\beta\gamma}$ is the totally anti-symmetric Levi-Civita tensor (see the supplemental material for details). 
In the following, we will parametrize the next-nearest neighbor hopping by the ratio $\kappa= \tilde \kappa/J_K$
as to keep the order of the overall energy scale $\tilde\kappa+J_K$ fixed.

Before we discuss this effective model two short remarks are in order. First, we note that the magnetic field term induces a second third-order term of the form $\sum_{j,k,l} \sigma_j^x \sigma_k^y \sigma_l^z$, where $j$, $k$, and $l$ are the three nearest neighbors of a common central site. This yields a local four-Majorana fermion interaction in the effective Hamiltonian. Closer inspection of this term shows that it is irrelevant in a renormalization group sense and we, therefore, neglect it in the following (see the supplementary material).
Second, we point out that tilting the magnetic field direction away from the 111 axis does not significantly alter the effective Hamiltonian. Such a tilt simply adds spatial anisotropies to the strength of the Kitaev interaction leading to a shift of the position of the Weyl points. Our results are in fact valid as long as the magnetic field has non-vanishing components along all three spatial directions.

\begin{figure}
  \includegraphics[width=\columnwidth]{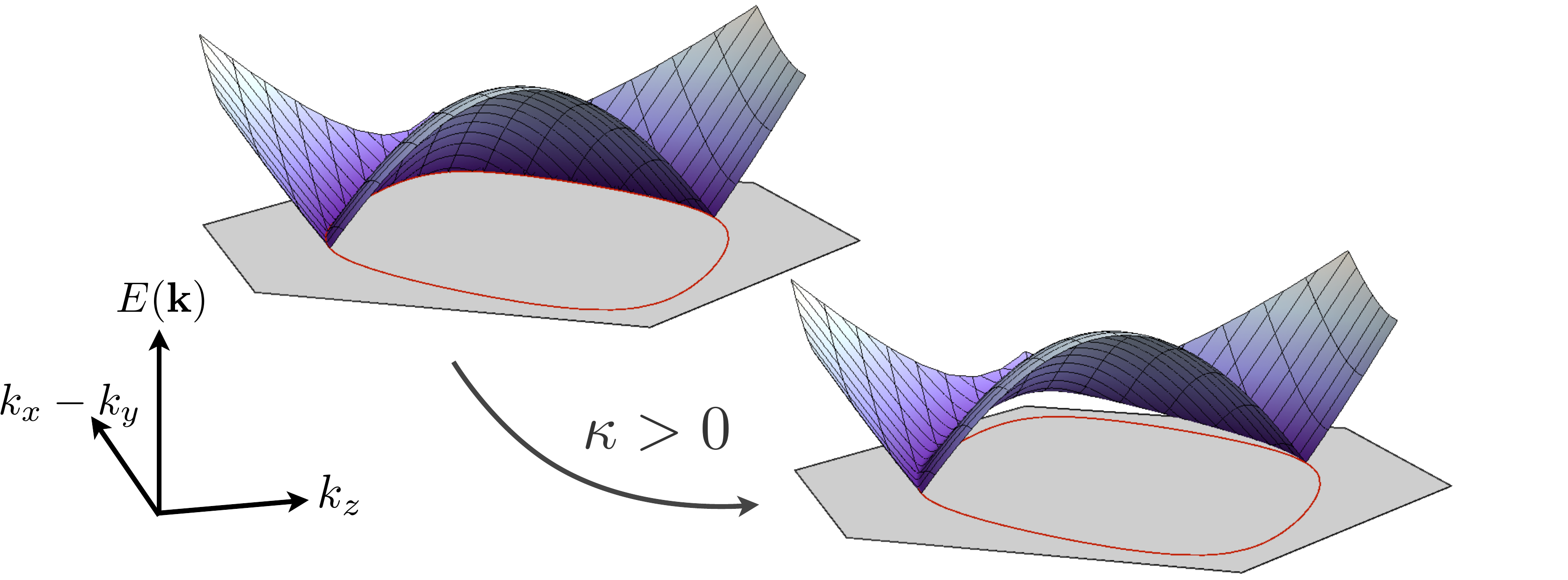}
  \caption{
    (color online) In the presence of a finite magnetic field along the 111 direction the Fermi line of the hyperhoneycomb Kitaev model is gapped out 
    except for two distinct Weyl points. 
   }
  \label{Fig:LineGapping}
\end{figure}

\noindent {\it Weyl points.--}
Returning  to the effective Hamiltonian \eqref{eq:eff}  
we find that a magnetic field immediately gaps out the gapless modes along the Fermi line -- except for two singular points. As illustrated in Fig.~\ref{Fig:LineGapping} we observe that the energy dispersion around these points is {\it linear} in all momentum directions and takes the form 
\[
	E(\mathbf q)=\sum_{i=1}^3 \mathbf  v_{i} \cdot \mathbf q \,\tau_i \,,
\] 
where the Pauli matrices $\tau_i $ act within the subspace spanned by the two touching bands (above and below the Fermi energy) and $\mathbf q$ is the momentum relative to the singular point \cite{FootnoteVelocities}. 
%
%
As such, these two remaining gapless points are in fact  a pair of Weyl points (WP). Such WPs have attracted considerable interest recently as it has been shown that these points describe {\em topologically protected} band touchings \cite{WeylSM}.
For electronic systems, the presence of such band touchings at the Fermi energy leads to a so-called Weyl semimetal -- a topologically protected gapless phase.
For the Majorana fermion system at hand, the WPs are fixed to exactly zero energy for symmetry reasons. Inversion symmetry assures that the WPs are at momenta $\mathbf Q$ and $-\mathbf Q$ and have identical energies $E(\mathbf Q )=E(-\mathbf Q)$, while particle-hole symmetry gives $E(\mathbf Q )=-E(-\mathbf Q)$. 

\begin{figure}[b]
  \includegraphics[width=\columnwidth]{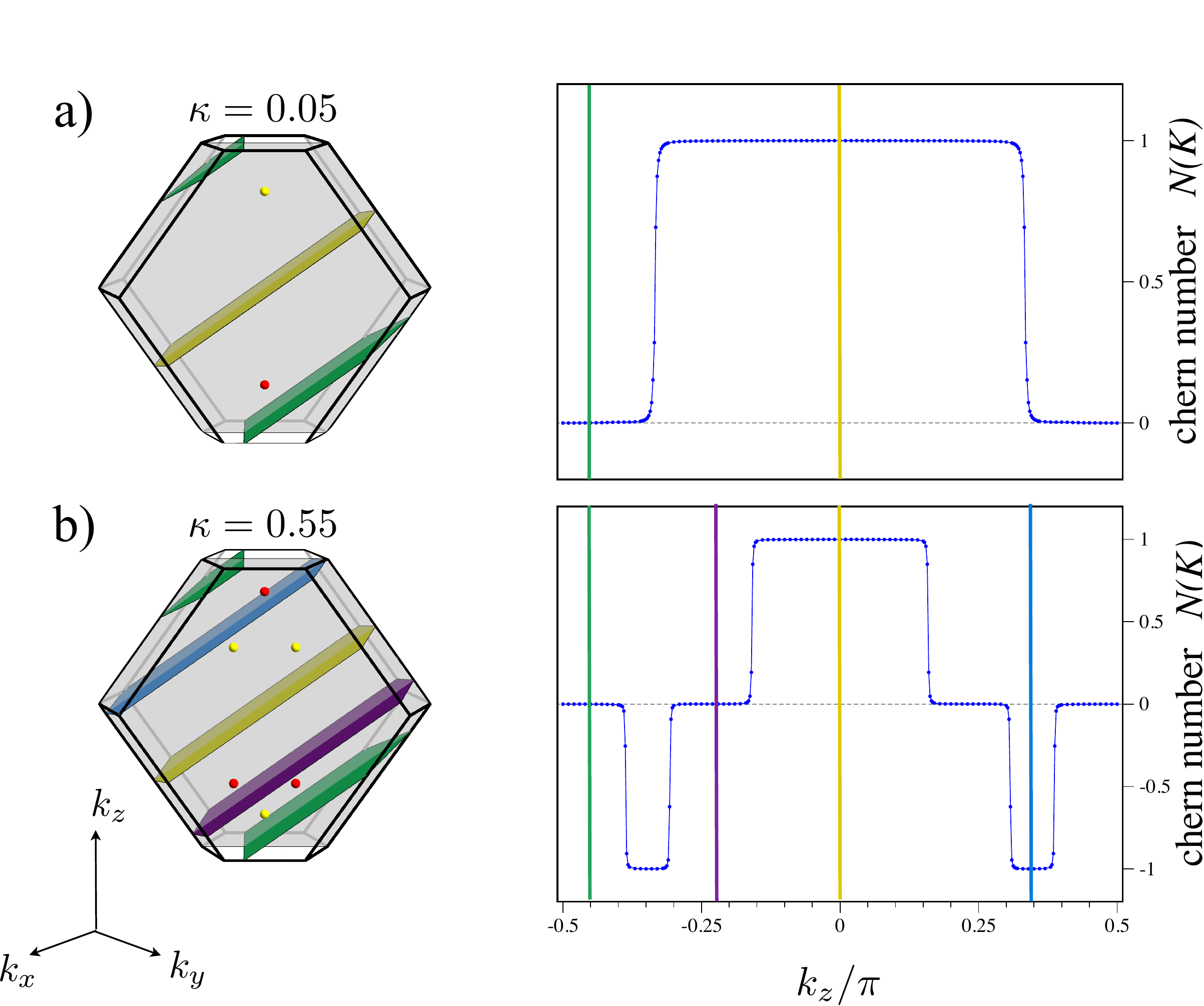}
   \caption{
    (color online) Plot of the Chern number of the effective Hamiltonian \eqref{eq:eff} restricted to the two-dimensional plane in reciprocal space defined by the three points $\mathbf k=(0,0,k_z)$, $\mathbf k+\mathbf q_2/2$ and $\mathbf k+\mathbf q_3/2$, where $\mathbf q_2$ and $\mathbf q_3$ are reciprocal lattice vectors (see the supplemental material for details). The colored lines indicate the position of the respective (hexagonal) planes shown on the left. 
   }
  \label{Fig:ChernNumbers}
\end{figure}

Each WP can be identified as a quantized source of Berry flux with the `charge' given by its chirality: sign$(\mathbf v_1 \cdot (\mathbf v_2 \times \mathbf v_3))$.  
One consequence of this is that the Chern number of the Hamiltonian restricted to a two-dimensional subspace of the Brillouin zone can be non-zero. To illustrate this effect, we calculate these Chern numbers on planes in momentum space as depicted in Fig.~\ref{Fig:ChernNumbers} where we parametrize the location of these planes by the momentum $k_z$. 
When passing through a WP the Chern number jumps by an amount related to the charge of the WP
\cite{FootnoteJump}.
This difference in the Chern number can be identified with the Chern number of a 2D surface surrounding the WP.
Thus, the WPs are indeed topological objects, explaining  their remarkable stability against any kind of local interaction -- 
WPs can only be gapped out in a pairwise fashion when two points of opposite chirality coincide at the same momentum. 

\begin{figure*}[t]
  \includegraphics[width= \linewidth]{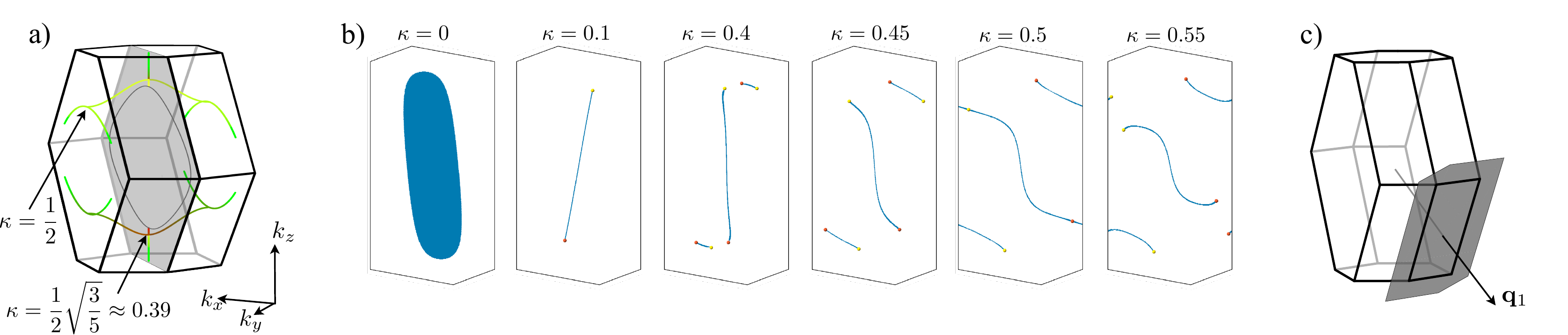}
  \caption{
    (color online)  Evolution of the a) Weyl points and b) corresponding Fermi arcs for increasing $\kappa$. The shading yellow-to-green indicates the evolution of negative chirality Weyl points for $\kappa=0 \rightarrow \infty$, shading red-to-green the evolution of their particle-hole partners with positive chirality (see main text). The surface Brillouin zone for breaking translation invariance in the $\mathbf a_1$-direction is indicated in c). 
   }
  \label{Fig:WeylPoints}
\end{figure*}

Intricately connected with the occurrence of non-trivial Chern numbers in the bulk is the presence of gapless surface states called Fermi arcs \cite{WeylSM}, which -- analogous to the bulk WPs -- are topologically protected. 
To see the emergence of such surface Fermi arcs in our spin model \eqref{eq:mag}, let us consider the effective Hamiltonian for a slab geometry, where periodic boundary conditions are imposed along the $\mathbf a_2$ and $\mathbf a_3$ directions, but not along $\mathbf a_1$ (see the supplemental material for a detailed description of the lattice).
The spectrum is then projected to the associated surface Brillouin zone, which is illustrated in Fig.~\ref{Fig:WeylPoints} c). For the time-reversal symmetric case ($\kappa=0$), the projection of the gapless Fermi line in the bulk is again a line filled with a flat surface band, shown in the left-most picture in Fig.~\ref{Fig:WeylPoints} b). 
Such flat surface bands in time-reversal symmetric models were first discussed in Refs.~\cite{SchnyderRyu,Burkov}. Their occurrence in three-dimensional Kitaev models was recently noted in Ref.~\cite{topologicalKim}.  
Breaking time-reversal symmetry ($\kappa\neq0$), the surface band develops a dispersion and only a single line connecting the projection of the two WPs remains at exactly zero-energy -- these are the Fermi arcs, which are illustrated in Fig.~\ref{Fig:WeylPoints} b) for various values of $\kappa$. 
We stress  that the WPs and their corresponding Fermi arc(s) are not protected by any symmetry, but rather by the topological nature of the WPs. Perturbing the system without annihilating  the WPs can only deform the Fermi arcs, but not destroy them. 
 
\noindent {\it Evolution of the Weyl points and Fermi arcs.--}
Let us now turn to a discussion of the effective Hamiltonian \eqref{eq:eff} for arbitrary $\kappa>0$. Notably, we find that the spectrum remains gapless for any value of $\kappa$, only the position and number of WPs change. 
The evolution of the WPs is shown in Fig.~\ref{Fig:WeylPoints} a) for $\kappa=0,\ldots, \infty$. 
In order to visualize the behavior for increasing $\kappa$, negative (positive) chirality WPs are shaded yellow (red) for $\kappa=0$ turning to green for $\kappa\rightarrow \infty$. 
Strikingly, at $\kappa=\frac 1 2 \sqrt{\frac 3 5 }$ each WP splits into three -- two of the same chirality and one of the opposite chirality such that the chirality is indeed  conserved locally. 
This behavior is also reflected in the evolution of the Fermi arcs. The single Fermi arc for small values of $\kappa$ splits into three separate fermi arcs precisely at $\kappa=\frac 1 2 \sqrt{\frac 3 5 }$. 
While the WPs on the high-symmetry line $\overline{\Gamma Y}$ recombine at $\kappa=\infty$, the WPs on the face of the Brillouin zone do not. 
Instead, the band gap collapses and the WPs merge into nodal lines that appear at $\kappa=\infty$. This behavior is exactly the opposite of the one shown in Fig.~\ref{Fig:LineGapping} that occurred when turning on $\kappa$. 
We should note here that this evolution for arbitrary $\kappa$ does not rigorously describe the physics at arbitrary magnet field strength $h$, as the perturbative expansion yielding the effective Hamiltonian \eqref{eq:eff} is, strictly speaking, only valid for small $\kappa\ll 1$, i.e. the limit in which no $\mathbb Z_2$ flux excitations are created.

\noindent {\it Thermodynamics.--}
Let us now reflect on what experimental probes may be used to detect a Weyl spin liquid. 
One possibility is to measure the different low-temperature contributions to the specific heat coming from the gapless bulk and surface modes. 
While the bulk contribution of the WPs to the specific heat results in a $T^3$-dependence on the temperature, the contribution from the Fermi arcs on the surface varies linearly \cite{FoonoteSpecificHeat} with $T$, i.e.
 \begin{align}
 C(T)\sim a_{\rm bulk} \cdot L^3 \cdot T^3 + a_{\rm surf} \cdot L^2 \cdot T \,,
 \label{eq:specificHeat}
 \end{align}
where $L$ is the linear system size and the prefactors $a_{\rm bulk / surf}$ are generically functions of $\kappa$ and depend on microscopic details. Varying the size and aspect ratios of samples, one can thereby  identify these two distinct distributions.
Measuring a specific heat of the form \eqref{eq:specificHeat} paradoxically indicating the presence of a {\em metallic} state in a Mott {\em insulator} is an unambiguous signature for the formation of a Weyl spin liquid.

Weyl spin liquids can also be probed via non-trivial transport features as they exhibit a thermal Hall effect. When applying a thermal gradient to the system, a net heat current {\em perpendicular} to the gradient arises due to the chiral nature of the surface modes.
This thermal Hall effect was first discussed in the context of Weyl superconductors \cite{WeylSuperconductors}, which are closely related to our system. Following the analysis of Ref.~\cite{WeylSuperconductors}, we can readily infer that a temperature bias in the $\hat x+\hat y$ direction in our system leads to a thermal Hall conductance $K$ (in the $\hat y-\hat x$ direction) that is proportional to the distance $d$ of the WPs in momentum space \cite{FootnoteThermalHallConductance}
\begin{align}
K =\frac 1 2  \frac{k_B^2\pi^2T}{3h} \frac{d}{2\pi}L_z \,,
\label{eq:thermalHall}
\end{align}
where $h$ is the Planck constant and $L_z$ is the length of the sample in the $z$ direction. 

\noindent {\it Discussion.--}
The hyperhoneycomb lattice is the first representative of an entire family of lattices, the so-called harmonic series of hyperhoneycomb lattices introduced in Ref.~\cite{hyperAnalytis}, which also reports the synthesis of the first-harmonic member of this family as a third crystalline form of Li$_2$IrO$_3$.
 The physics of the Kitaev model is very similar for all members of this harmonic series. In the presence of time-reversal symmetry the low-energy gapless modes of all of these model variants form a Fermi line \cite{topologicalKim}. As such, we also expect  similar behavior when breaking time-reversal symmetry, i.e. all members of this family will exhibit a Weyl spin liquid with all of the aforementioned properties and experimental signatures. 
%
%
The occurrence of Weyl spin liquids in this family of hyperhoneycomb lattices should be contrasted to the physics of the Kitaev model on the hyperoctagon lattice of Ref.~\onlinecite{hyperoctagon} (and its higher harmonics), where the gapless modes form a Fermi surface. Breaking time-reversal symmetry for these models does not destroy the Fermi surface, but merely deforms it. In fact, this deformation stabilizes the spin liquid ground state as it removes  possible BCS pairing instabilities \cite{BCS}. 

A more comprehensive picture for the emergence of Weyl spin liquids in three-dimensional Kitaev-type models arises if one frames the symmetries of the spin system and its underlying Majorana model in terms of the symmetry classification scheme of free fermion systems \cite{AltlandZirnbauer}. The situation described here -- particle-hole symmetry plus broken time-reversal symmetry -- corresponds to symmetry class D. In this class, WPs appear generically at zero-energy, if inversion symmetry is {\em not} broken. 
Only the location and number of WPs depend on microscopic details \cite{FootnoteWeylPoints}.
It should be noted that in the reverse situation of preserving time-reversal symmetry and breaking inversion -- corresponding to symmetry class BDI -- a Kitaev model cannot harbor a Weyl spin liquid.
The latter is due to the fact that particle-hole symmetry entails that WPs at momenta $\pm \mathbf Q$ have opposite chirality \cite{NielsenNinomiya}, while time-reversal symmetry restricts them to have the same chirality. Thus, WPs of opposite chirality must necessarily coincide and even infinitesimal perturbations can gap them out pairwise
\cite{FootnoteSymmetry}. 
This should be contrasted to electronic systems where one can find WPs in systems that break {\em either} time-reversal or inversion symmetry, corresponding to symmetry classes A or AII, respectively -- see Refs. \cite{Burkov} and \cite{Murakami} for examples of minimal electronic models realizing these symmetry classes. 
While disorder effects for topological insulators in all free-fermion symmetry classes are well understood \cite{TI-tenfold}, such a comprehensive picture does not yet exist for topological semimetals. It would be interesting to study whether Weyl spin liquids (symmetry class D) exhibit different disorder effects than Weyl semimetals in electronic systems (symmetry class A or AII).

Finally, taking a step back we note that our motivation to study such three-dimensional generalizations of the Kitaev model arises from a {\em strong-coupling} perspective of spin-orbit entangled $j=1/2$ Mott insulators found in a number of Iridates in close proximity to a metal-insulator transition. 
It is quite satisfying to see that these models are capable of capturing the emergence of a Weyl spin liquid -- a state, in which the collective physics of the localized, spin-orbit entangled degrees of freedoms of a weak Mott insulator closely mimics the itinerant electronic state of a nearby Weyl semimetal. This Weyl semimetal has been found when studying these materials from the opposite limit of a {\em weak-coupling} perspective \cite{WeylSM}.

\noindent {\it Acknowledgments.--}
We thank A. Akhmerov, T. Hughes, A. Rosch and especially S. Bhattacharjee and Y. B. Kim for insightful discussions. We acknowledge partial support from SFB TR 12 of the DFG.
The numerical simulations were performed on the CHEOPS cluster at RRZK Cologne.



\begin{thebibliography}{00}
\bibitem{footnoteZeroFlux} 
Numerical studies done independently by ourselves and previously in Ref.~\cite{Mandal09} indicate that the ground state lies in the flux-free sector -- analog to the two-dimensional case of the honeycomb lattice, for which this result is rigorously known due to a theorem by Lieb \cite{Lieb}. While Lieb's theorem does generalize to certain three-dimensional lattice geometries, it cannot be readily applied to the hyperhoneycomb lattice.
\bibitem{FootnoteVelocities}
This equation also defines the three velocities $\mathbf v_i$ of the gapless modes emanating from the WPs. Their precise numerical value can be computed from the numerically obtained band structure. For our argument their only relevant property is that they are linearly independent, see Ref.~\cite{WeylSM}.
\bibitem{FootnoteJump}
We note that the system always remains globally neutral with respect to the Weyl charge.
\bibitem{FoonoteSpecificHeat}
Our argument assigns the Fermi arcs purely to the surface. However, since the Fermi arcs penetrate into the bulk in the vicinity of the projected WPs, one could imagine additional contributions to the specific heat which give rise to an intermediate scaling regime (with intermediary temperature and system size scaling exponents).
\bibitem{FootnoteThermalHallConductance}
This formula applies strictly speaking only for small magnetic fields, when there is only a single, straight Fermi arc connecting the two WPs.
\bibitem{FootnoteSymmetry}
This argument can in fact be generalized to show that all chiral symmetry classes cannot realize a Weyl semimetal \cite{WeylSuperconductors2}.
\end{thebibliography}
\end{document}